\documentclass[%
 aip,
 sd,%
 amsmath,amssymb,
preprint,%
]{revtex4-1}

\usepackage{graphicx}
\usepackage{dcolumn}
\usepackage{bm}

\usepackage{setspace,color,url}
\begin{document}

\preprint{Accepted manuscript, not for distribution}

\title{Molecular dynamics simulation of the electrokinetic flow of an aqueous electrolyte solution in nanochannels}
%
\author{Hiroaki Yoshida}
\email{h-yoshida@mosk.tytlabs.co.jp}
\affiliation{Toyota Central R\&D Labs., Inc., Nagakute, Aichi
 480-1192, Japan}
\affiliation{Elements Strategy Initiative for Catalysts and Batteries
(ESICB), Kyoto University, Katsura, Kyoto 615-8520, Japan}
\author{Hideyuki Mizuno}
\affiliation{Laboratory for Interdisciplinary Physics, UMR 5588,
Universit\'e Grenoble 1 and CNRS, 38402 Saint Martin d'H\`eres, France}
\author{Tomoyuki Kinjo}
\author{Hitoshi Washizu}
\affiliation{Toyota Central R\&D Labs., Inc., Nagakute, Aichi
 480-1192, Japan}
\affiliation{Elements Strategy Initiative for Catalysts and Batteries
(ESICB), Kyoto University, Katsura, Kyoto 615-8520, Japan}
\author{Jean-Louis Barrat}%
\affiliation{Laboratory for Interdisciplinary Physics, UMR 5588,
Universit\'e Grenoble 1 and CNRS, 38402 Saint Martin d'H\`eres, France}
\affiliation{Institut Laue--Langevin, 6 rue Jules Horowitz, BP 156, 38042 Grenoble, France}
\date{\today}
%
\begin{abstract}
Electrokinetic flows of an aqueous NaCl solution in nanochannels with
negatively charged surfaces are studied using  molecular dynamics (MD) simulations.
The four  transport coefficients that characterize the response to
weak electric and pressure fields,
namely the
coefficients for the electrical current in response to the electric field
($M^\mathrm{jj}$) and the pressure field ($M^\mathrm{jm}$), and those for the mass flow in
response to the same fields ($M^\mathrm{mj}$ and $M^\mathrm{mm}$),
are obtained in the linear regime using a Green--Kubo approach.  
Nonequilibrium simulations with explicit external fields are also carried out, and 
the current and mass flows are directly obtained.
The two methods exhibit good agreement
even for large external field strengths, and 
Onsager's reciprocal relation ($M^\mathrm{jm}=M^\mathrm{mj}$) is  numerically confirmed in both approaches.
The influence of the surface charge density on the flow is also considered.
The values of the transport 
coefficients are found to be smaller for  larger surface charge density, because 
the counter-ions strongly bound near the channel surface interfere 
with the charge and mass flows. 
A reversal of the streaming current 
and of the reciprocal electro-osmotic flow, with a change of sign of $M^\mathrm{mj}$ due to the excess co-ions, takes places for very
high surface charge density.
\end{abstract}
\maketitle
%
%
\section{\label{sec_intro}Introduction}
Power generation and energy storage technology 
that utilize electrochemical devices, such as the
lithium-ion battery and the fuel cells, have been studied extensively,
and their performance has continued to improve regularly.
At the core of these electrochemical devices, one generally finds
 systems consisting of complex electrolyte solutions
and   of charged solids, e.g.,
the porous electrode layer in the lithium-ion battery~\cite{SG2010}
and the electrolyte membrane in the fuel cell.~\cite{PRA2010}
The transport of the ions and the solvent
through the charged solid structure 
affects the total performance of the devices significantly,
and thus the control and optimization of the
transport phenomena are central areas of research in the
development of  innovative electrochemical devices.~\cite{KDS2009,SPB+2013,BYN+2014}

Transport phenomena are observed as
flows of the ions and solvent in response to external driving forces.
The driving forces important in electrochemical systems are 
those induced by the electric field and/or the pressure gradient.~\cite{N1991}
If the system is close to the thermal equilibrium state
so that the system responds linearly to the external fields,
the electric current ${J}$ and mass flow ${Q}$ 
induced by the electric field and the pressure gradient are written in the following form:~\cite{BA2004a}
\begin{equation}
\left(\begin{array}{c}
{J} \\ {Q}
			\end{array}\right) 
=
\left(\begin{array}{cc}
M^{\mathrm{jj}} & M^{\mathrm{jm}}\\
M^{\mathrm{mj}} & M^{\mathrm{mm}}
		 \end{array}\right)
\left(\begin{array}{c}
 E_x \\ P_x
\end{array}\right),
\label{s1-mat}
\end{equation}
where $E_x$ is the electric field and $P_x$ is the mass acceleration
representing the pressure gradient.~\footnote{
{
The flux of one ion component (cation or anion)
is another important response of the present system,
and then the corresponding external field is the chemical potential gradient
of that component;
although in the present paper we restrict ourselves
to investigating two fluxes, i.e., the current and mass flow,
to cover the transport coefficients of $3\times 3$ matrix 
is a possible extension of this work.}}
The coefficient $M^{\mathrm{jj}}$ corresponds to the effective electric conductivity,
and $M^{\mathrm{mm}}$ is directly related 
to the permeability of the porous media.~\cite{Bear1988}
On the other hand, 
$M^{\mathrm{jm}}$ and $M^{\mathrm{mj}}$ are the transport coefficients
for the streaming current and the electro-osmotic flow.
Onsager's reciprocal relation states that the values of these
coefficients
are identical, i.e., $M^{\mathrm{jm}}=M^{\mathrm{mj}}$ 
(see e.g., Refs~\onlinecite{O1931A,*O1931B,DM1962,HM2006,BA2004a,BC2010B}).
The values of the transport coefficients are dependent
on complex, multi-physics phenomena,
namely the internal and external electric field, 
the solvent flow, and the diffusion and migration of ions.
Furthermore, commonly used electrochemical systems exhibit a broad hierarchy of scales:
whereas the atomic scale is important at the 
interface between the solid and the electrolyte solution,
the thickness of the electrical double layer formed near the interface
can extend to a few tens of nanometers,
and the characteristic size in the porous media or membranes 
through which the electrolyte solution
flows ranges from a few nanometers to tens of micrometers.
It is therefore very difficult to 
evaluate the values of the transport coefficients 
using a precise model that describes all of the 
physics included in the system.
One strategy to overcome this difficulty is
to incorporate the effect of the interface
as boundary conditions for the macroscopic description
with coarse-graining the events near the interface,
and to evaluate the transport coefficients
using  equations based on the continuum theory.~\cite{BA2004a,LC2006,ABD+2013,ODJ+2013}
Obviously, however, the macroscopic description is not always valid. 
When the  relevant length scales decrease, the relative effect in the atomic scale becomes
significant, and the theoretical predictions fail to reproduce the
experimental observation.~\cite{KBA2005,ZHC2003}
In this case, since the characteristic scale approaches the atomic scale
and the scale gap is less important,
the molecular dynamics (MD) method,
which deals explicitly with atoms,
becomes  accurate and efficient.~\cite{RP2013}
Applying realistic interaction forces
between particles constituting ions, solvent molecules, and
charged solids allows one to evaluate the transport coefficients
by capturing accurately the phenomena taking place at interfaces.

Two methods are available to obtain 
the responses (current ${J}$ and the mass flow ${Q}$) to external fields 
using MD simulations.
One is to assume linear response, and to apply the Green--Kubo formulas.~\cite{BB1994,MDJ+2003,HM2006,BB2013}
The four transport coefficients are then obtained
simultaneously 
using the results of  MD simulations performed 
at thermal equilibrium, without any external force.
Although the Green--Kubo formulas are only valid within the 
limit of the linear response regime,
the real systems under usual conditions are most often operating within this limit,
in view of the fact that the current and mass flow observed experimentally respond linearly
to the external fields (e.g., Refs.~\onlinecite{MEM+1979,SYZ+2004}).
{
(For systems in which
a non-linear response
is important, the transient time correlation function formalism is
a possible alternative to the Green--Kubo
approach.~\cite{EM2008,BBS+2012})}
The other method is
to directly observe the current and mass flow, after 
applying  an explicit external field in the MD
simulation.~\cite{QA2004,HCY+2008,HCY+2008a,LWCR2010,BMR+2013}
This method is referred to as the direct method in the present paper.
Since the direct method does not assume the system
to be close to thermal equilibrium, 
the charge and mass flow can be obtained  in response to
external fields of arbitrary strength.
Further, since the actual flows are induced in the simulations,
detailed discussion on the profiles of flows is possible.
However, 
the field strength attainable in laboratories 
is too small to distinguish the induced flows from the thermal fluctuation,
and extremely strong external fields are necessary.
Therefore, comparison with the results based on the linear response
theory is inevitable, before extrapolating the results of the direct
method to real systems.

In the present study,
we apply the two methods 
to specific systems with 
realistic ions and solvent.
More precisely, 
the behavior of an aqueous NaCl solution 
in a channel between two charged walls
is investigated using MD simulations.
The material of the wall is not specified,
and simply corresponds to a generic hydrophilic material. 
Although the
results of each method for 
similar systems have been reported,~\cite{MDJ+2003,QA2004,HCY+2008,BMR+2013} 
a systematic comparison of the 
transport coefficients obtained
using the Green--Kubo formulas
with the results of the direct method is,
to our knowledge, new.
Here, we establish the 
protocol to evaluate the transport coefficients
using the data from MD simulations,
and we compare 
systematically
the current and mass flow
obtained through~Eq.~\eqref{s1-mat}
using the transport coefficients
with those obtained 
by using the direct method.
In addition,
the influence of the surface, which is important for  the flow at a small scale, is investigated.
Special attention is devoted to  the effect of the surface charge density, because
the counter-ion condensation at the charged surface
must have important effects on the response.~\cite{WK2002,Manning2011}
The dependence of the effective electrical conductivity
and of the flow rate  in a Poiseuille  geometry is examined,
and the inversion of the streaming current and the electro-osmotic
flow~\cite{QA2004} is discussed.

%
%
\section{\label{sec_model}Models and methods}
\begin{figure}[t]
\begin{center}
\centering
\includegraphics[scale=1.1]{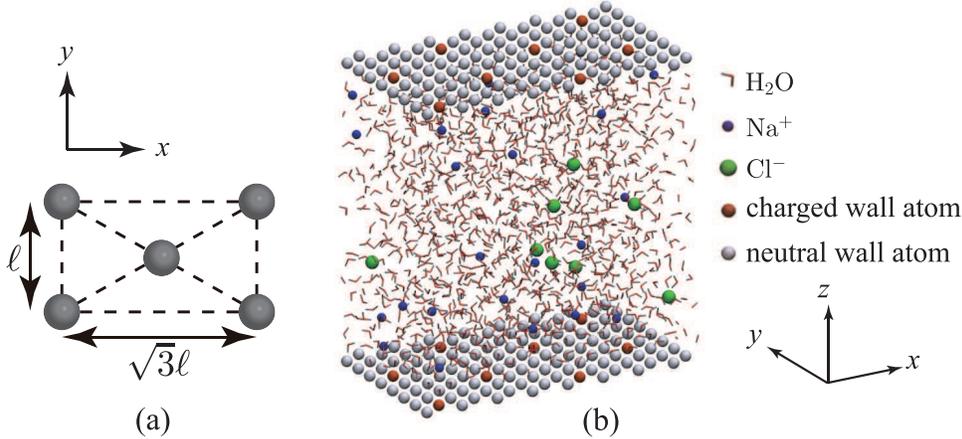} 
\caption{
(a) A unit cell of the equilateral triangular lattice of the wall atoms in the $x$-$y$ plane.
(b) A snapshot of the system at equilibrium.
}
\label{fig01}
\end{center}
\end{figure}

We consider an aqueous NaCl solution 
between two charged walls, as shown in Fig.~\ref{fig01}.
Each wall consists of 
two-dimensional equilateral triangular lattice
of a model atom, say atom A,
with shortest distance between two atoms being $\ell$.
In the present study, $\ell$ is fixed at $3$\,\AA.
Among the wall atoms, 
every $\ell_c/\ell$ atoms in the direction of the shortest distance
are negatively charged with one elementary charge ($-e$).
The absolute value of the surface charge density is then 
expressed as $\sigma=2e/\ell_c^2\sqrt{3}$. 
Since in the present study the walls are always negatively charged,
the surface charge density given in the following refers to its absolute magnitude.
The electrolyte solution contains
$N^{\mathrm{Cl}}$ Cl$^-$ ions and $N^{\mathrm{Na}}$ Na$^+$ ions.
The relation $N^{\mathrm{Na}}=N^{\mathrm{Cl}}+N^{\mathrm{Ac}}$
holds with $N^{\mathrm{Ac}}$ being the number of charged atom A,
because of the electrical neutrality.

The extended simple point charge (SPC/E) model~\cite{BGS1987} is employed
to describe  the  interactions between water molecules.
The interactions between ions are
described simply by a sum of electrostatic and Lennard-Jones (LJ) potentials,
with parameters  taken from Ref.~\onlinecite{SD1994}.
The LJ parameters for water-ion and Na-Cl
pairs are determined by the 
Lorentz--Berthelot mixing rule.~\cite{AT1989,HM2006}
For the interaction between a wall atom A
and a water molecule, 
we employ the following potential $\varphi$:~\cite{PG2008}
\begin{equation}
\varphi(r)=
\dfrac{\eta\epsilon}{n-m}
\left(m\left(\dfrac{r_0}{r}\right)^n
-n\left(\dfrac{r_0}{r}\right)^m
\right), 
\label{s2-aho}
\end{equation}
where $r$ is the distance between atoms, and 
$r_0$ is the distance at which $\varphi/\eta$ takes the minimum value $-\epsilon$\,;
$m$ and $n$ are integers.
The values of the parameters used in the simulation
are listed in Table~\ref{table01}.
The factor $\eta=2.34$
common to A-Hydrogen and A-Oxygen interactions
was determined
such that the
binding energy between an SPC/E
molecule and the triangular lattice of atom A
was equal to that of the lowest energy
between two SPC/E molecules ($-7.5$\,kcal/mol).
{
Then, the model surface employed here
represents a hydrophilic surface
within the limitation of homogeneously distributed sites.
Note that in real systems the
hydrophilic sites are distributed more heterogeneously.~\cite{SMC+2013}}
In calculating the LJ interaction forces between a wall atom A and an ion,
the mixing rule mentioned above is employed, 
with the LJ parameters of the neutral and charged wall atoms
being the same as those of the SPC/E model and the Cl$^-$ ion, respectively.

\begin{table}[t]
\vspace*{0cm}
\vspace*{0cm}
\centering
\caption{Parameters for interaction between atom A and a water molecule.}
\label{table01}
\vspace*{0mm}
\renewcommand{\arraystretch}{1.2}
\begin{tabular}{cccccccccc}
\hline
\hline
interaction & $m$ & $n$ & $r_0$\,[\AA] & $\epsilon$\,[kcal/mol] & $\eta$ \\
\hline
A--O & 6 & 12 & 3.85 & 0.25 & 2.34\\
A--H & 8 & 12 & 2.14 & 1.52 & 2.34\\
\hline
\hline
\end{tabular}
\end{table}

The MD simulations are carried out
using the open-source code LAMMPS.~\cite{LAMMPS,Plimpton1995}
During the simulations, the number of particles
and the volume $V$ are kept constant while 
the temperature $T=300$\,K is maintained using
the Nos\'e--Hoover thermostat (NVT ensemble).
The time step is $1$\,fs throughout this paper,
{
with using SHAKE algorithm~\cite{RCB1977} 
to maintain the SPC/E water molecules rigid.}
The LJ interactions are treated using the standard method of
spherical cutoff (cutoff radius $=9.8$\,\AA),
while long-range Coulomb interactions are treated by using 
the particle-particle particle-mesh (PPPM) method.
In order to treat the slab geometry, the method proposed by Yeh and Berkowitz \cite{YB1999} is
employed, i.e., 
the $z$ direction is first extended to create empty spaces outside the channel, then the periodic boundary
conditions are applied; 
the artifacts from the image charges due to periodic boundary
conditions in the $z$ direction are removed by adding a correction force to each particle.

The distance $H$ between the upper and lower walls
is determined such that the normal  pressure is equal to the atmospheric pressure,
in the following manner:
first the water molecules and the ions
are randomly distributed at a density lower than 
that at atmospheric pressure,
and an MD simulation is 
carried out with this configuration as an initial condition.
During the simulation, the atoms of the upper wall
are constrained such that they move only in the $z$ direction,
while the atoms of the lower wall are completely frozen.
At each time step, 
the forces in the $z$ direction
felt by all of the upper-wall atoms are averaged (denoted by $\bar{f}_{wz}$).
Then the forces acting on the upper-wall atoms 
are replaced by the common force $\bar{f}_{wz}-f_0$,
with $f_0$ being the force per atom corresponding to the atmospheric pressure.
{
Typically it takes $10$\,ps for $\bar{f}_{wz}-f_0$ to reach zero,
after which it fluctuates.} 
The simulation is continued for $0.6$\,ns,
and the average value of the distance between the upper and lower
walls, 
{
over the interval $0.1 < t < 0.6$\,ns},
is chosen as $H$.
The configuration obtained after $1$\,ns equilibration time,
with the upper- and lower-wall atoms being frozen
at a distance $H$, is used as the 
initial condition for the following simulations.

In the present study,  MD simulations under explicit external
fields (direct method), as well as  equilibrium simulations, are
carried out. 
The forces acting on $i$th particle due to the external fields are given by
\begin{align}
F^E_{xi}=q_i E_x,\\
F^P_{xi}=m_i P_x,
\end{align}
where $q_i$ and $m_i$ are respectively the charge and the mass of $i$th particle,
$E_x$ is the electric field in the $x$ direction,
and $P_x$ is the mass acceleration for simulating the force due to the pressure
gradient while applying  periodic boundary conditions.
The relation between $P_x$ and the pressure field $p$ is 
$P_x=-(1/\rho_0)(\mathrm{d}p/\mathrm{d}x)$ ($\rho_0$: the average density).
The responses to the fields, namely the charge flux $j_x$ and the mass flux $c_x$, 
are obtained from the MD trajectory as
\begin{align}
& c_x =\sum_{i} m_i \dot{x}_i,
\label{s2-mflux}\\
& j_x =\sum_{i} q_i \dot{x}_i,
\label{s2-qflux}
\end{align}
where the summation runs over all particles.

%
%
\section{\label{sec_gk}The Green--Kubo formulas}

The current density ${J}=j_x/V$ and the mass flow 
density ${Q}=c_x/V$
under the fields $E_x$ and $P_x$ are
obtained using Eq.~\eqref{s1-mat}
for the system  within the limit of the linear response regime,
as described in Introduction.
The transport coefficients 
$M^{\mathrm{jj}}$, $M^{\mathrm{jm}}$, $M^{\mathrm{mj}}$, and
$M^{\mathrm{mm}}$ are
related to the time-correlation functions
of the charge and mass fluxes via the 
Green--Kubo formulas based on the linear response theory.

In order to derive the specific forms of the Green--Kubo formulas
for the system considered in the present study,
we follow the standard discussion of the linear response theory.~\cite{HM2006}
The Hamiltonian of the system $\mathcal{H}$
under a weak external field $\mathcal{F}_0$ is
perturbed by $\mathcal{H}'(t)$ from the
value at the thermal equilibrium state
$\mathcal{H}_{\mathrm{eq}}$:
\begin{align}
&\mathcal{H}=\mathcal{H}_{\mathrm{eq}}+\mathcal{H}'(t),
\label{s3-hamiltonian}\\
&\mathcal{H}'(t)=-\mathcal{A}(\bm{r}^N)\mathcal{F}_0
\exp\left(-i\omega t\right),
\label{s3-yuragi}
\end{align}
where $\omega$ is the frequency of the external field,
and $\mathcal{A}$ is a function of the particle positions $\bm{r}^N$.
Then, the change in the observed variable $\mathcal{B}$, denoted by
$\Delta\mathcal{B}$, is expressed as
\begin{align}
&\langle\Delta\mathcal{B}\rangle
=
M^{BA}(\omega)\mathcal{F}_0\exp(-i\omega t),
\label{s3-cb}\\
&M^{BA}(\omega)
=\dfrac{1}{k_{\mathrm{B}} T}\int_0^{\infty}
\langle\mathcal{B}(t)\dot{\mathcal{A}}\rangle\exp(i\omega t)
\mathrm{d}t,
\label{s3-ggk}
\end{align}
where $k_\mathrm{B}$ is the Boltzmann constant.
In the system considered herein,
$\mathcal{F}_0=E_x$ and $\mathcal{A}=\sum_i q_ix_i$ in the case of
electric field,
and $\mathcal{F}_0=P_x$ and $\mathcal{A}=\sum_im_i x_i$ in the case of
mass acceleration.
The observed variables are
the current density $\mathcal{B}={J}=j_x/V$,
and the mass flow density 
$\mathcal{B}={Q}=c_x/V$.
Since a time-independent external field is considered ($\omega\to 0$),
the transport coefficients are expressed
in terms of $j_x$ and $c_x$ as follows:
\begin{align}
 &M^{\mathrm{jj}}
=\dfrac{1}{k_{\mathrm{B}}TV}\int_0^{\infty}
\langle j_x(t)j_x(0)\rangle\mathrm{d}t,
\label{s3-cqq}\\
 &M^{\mathrm{jm}}
=\dfrac{1}{k_{\mathrm{B}}TV}\int_0^{\infty}
\langle j_x(t)c_x(0)\rangle\mathrm{d}t,
\label{s3-cqm}\\
 &M^{\mathrm{mj}}
=\dfrac{1}{k_{\mathrm{B}}TV}\int_0^{\infty}
\langle c_x(t)j_x(0)\rangle\mathrm{d}t,
\label{s3-cmq}\\
 &M^{\mathrm{mm}}
=\dfrac{1}{k_{\mathrm{B}}TV}\int_0^{\infty}
\langle c_x(t)c_x(0)\rangle\mathrm{d}t.
\label{s3-cmm}
\end{align}
Here, definition of $M^{\mathrm{jj}}$ is identical to that of the
electrical conductivity, 
and $M^{\mathrm{mm}}$ is related 
to the permeability of the porous media
: $k=\nu M^{\mathrm{mm}}/\rho_0$ 
($k$: permeability, $\nu$: kinetic viscosity).~\cite{Bear1988}
The coefficients $M^{\mathrm{jm}}$ and $M^{\mathrm{mj}}$
are the measures of the streaming current and the electro-osmotic flow,
respectively.
Since the MD simulation is 
time reversible apart from the numerical error,
Eqs.~\eqref{s3-cqm} and \eqref{s3-cmq} are identical
in the thermal equilibrium state, i.e.,
$M^{\mathrm{jm}}=M^{\mathrm{mj}}$,
which is known as Onsager's reciprocal relation.~\cite{O1931A,*O1931B,DM1962,BA2004a,HM2006,BC2010B}
\begin{figure}[t]
\begin{center}
\includegraphics[scale=1]{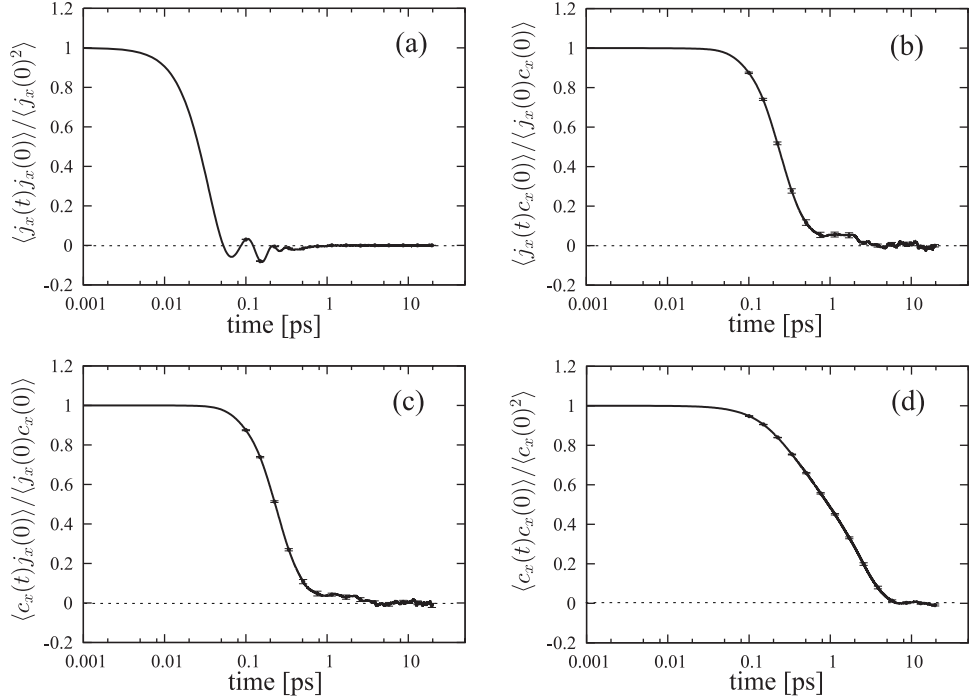} 
\caption{Normalized time-correlation functions of the charge and mass
 fluxes:
(a) charge-charge, (b) charge-mass, (c) mass-charge, and (d) mass-mass.
Each curve is the average
over ten simulation runs with different
initial configurations, and the standard error is indicated
with the error bar.
}
\label{fig02}
\end{center}
\end{figure}  
%

%
%
\section{\label{sec_result}Results and discussion}

%
\subsection{\label{sec_first}Evaluation of the transport coefficients}
We first apply the 
Green--Kubo formulas~\eqref{s3-cqq} through \eqref{s3-cmm}
to a specific system with an NaCl solution confined in
a nanochannel.
Here we consider 
the walls with $8$ units cells both in the $x$- and $y$-directions.
(A unit cell is shown in Fig.~\ref{fig01}(a).)
Every four atoms in the direction of the shortest distance of the
triangular lattice are negatively charged ($-e$).
The surface charge density is then $\sigma=0.128$\,C/m$^2$.
The gap between the walls contains $1260$ water molecules
with $24$ Na$^+$ ions and $8$ Cl$^-$ ions.
The charged wall atoms and the ions maintain the electrical neutrality.
The distance between the upper and lower walls
determined with the strategy described in 
Section~\ref{sec_model}
is $H=41.1$\,{\AA}, and the resulting molar concentrations of Na$^+$ and
Cl$^-$ are $0.97$\,M and $0.32$\,M, respectively.
\begin{figure}[t]
\begin{center}
\includegraphics[scale=1]{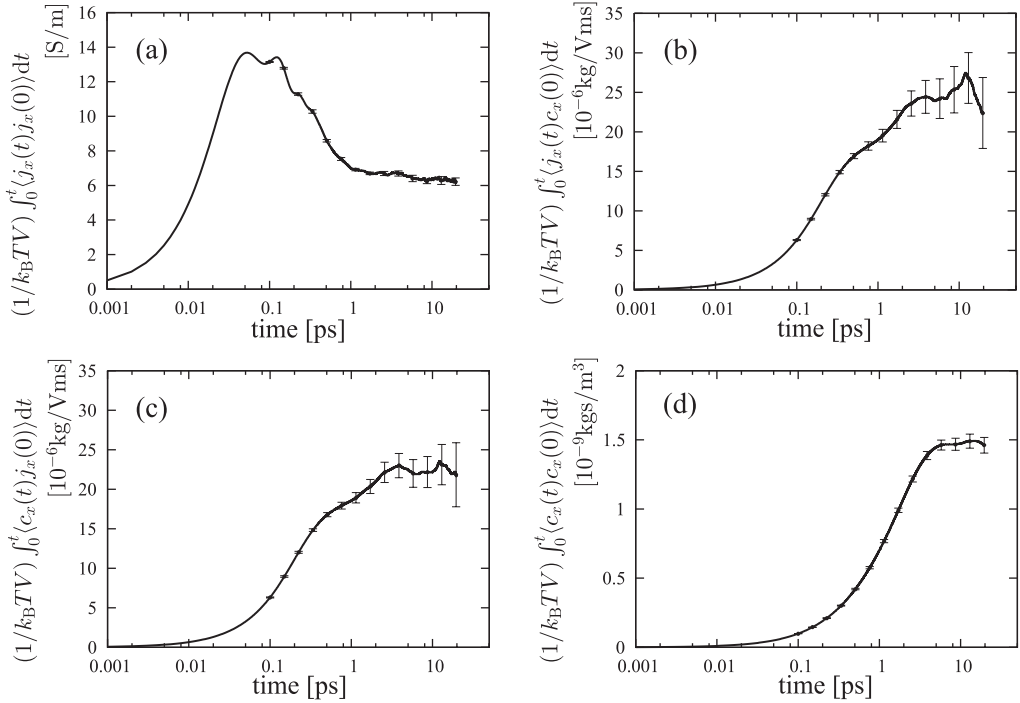} 
\caption{Time-integrated correlation functions of the charge and mass
 fluxes:
(a) charge-charge, (b) charge-mass, (c) mass-charge, and (d) mass-mass.
The values are scaled by $k_{\mathrm{B}}TV$, so that
their long-time limits directly correspond to the transport coefficients.
See the caption of Fig.~\ref{fig02} for the meaning of the error bar.
}
\label{fig03}
\end{center}
\end{figure}  
\begin{figure}[t]
\begin{center}
\includegraphics[scale=1]{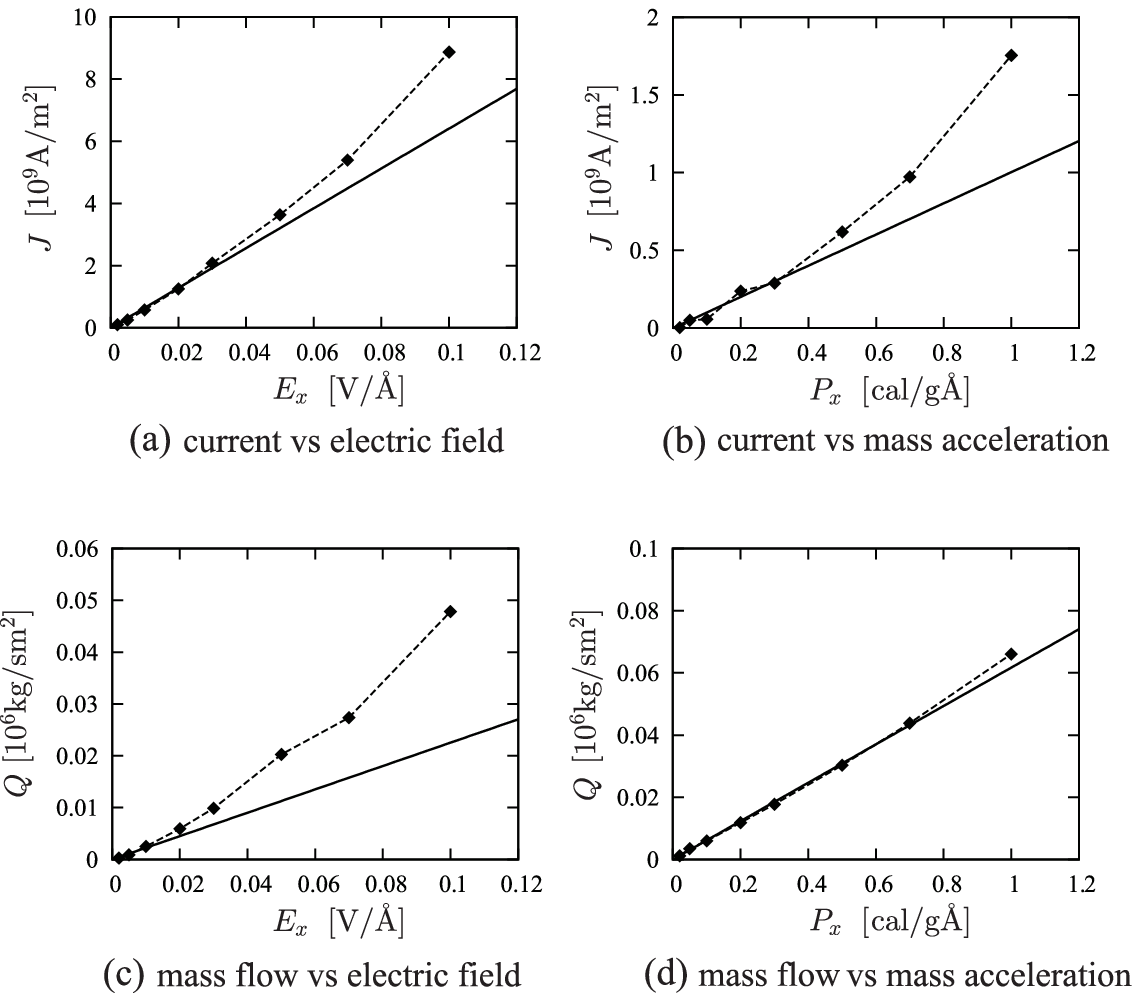} 
\caption{
Current and mass flow densities induced by the electric field or the mass
 acceleration.
The symbol indicates the results of the simulations with the
explicit external field, and the solid linear line 
indicates Eq.~\eqref{s1-mat} with the transport coefficients 
obtained from the Green--Kubo formulas.
}
\label{fig04}
\end{center}
\end{figure}  

In order to obtain the time-correlation functions 
of $j_x$ and $c_x$ necessary for Eqs.~\eqref{s3-cqq} through \eqref{s3-cmm},
the MD simulation at  thermal equilibrium is performed for $5$\,ns, and the values of $j_x$ and $c_x$ are recorded every time step ($1$\,fs).
The correlations are taken for the time difference $0\le t\le 20$\,ps,
and the $498\times 10^{4}$ time-series samples are averaged.
We have checked the influence of the initial configuration, and have observed
considerable variations in the time-correlation functions,
especially in $\langle j_x(t)j_x(0)\rangle$, $\langle j_x(t)c_x(0)\rangle$, and $\langle c_x(t)j_x(0)\rangle$.
The variation due to the initial configuration was 
not suppressed  even if we extended the simulation time to
$20$\,ns and increased the number of the time-series samples.
This is because the motion of ions in the electrolyte solution 
is slow (cf. Ref.~\onlinecite{LL1996}), 
and very long simulation time is required to
obtain the sufficient statistics in calculating the time-correlation functions in which the motion of the ions is important.
In the present study, we circumvent this difficulty
by carrying out ten MD simulations with different initial configurations,
each of which runs for $5$\,ns.
The ten time-averaged correlation functions are then averaged
again, and the results are shown in Fig.~\ref{fig02}.
The error bar for ten simulation runs shows that the
error due to a specific initial condition is greatly reduced.
As shown in Fig.~\ref{fig03}, however,
if the correlation functions are integrated over very long time,
the error accumulates and becomes significant.
Therefore, the time-integration
in Eq.~\eqref{s3-cmm} is terminated at $10$\,ps,
and that in Eqs.~\eqref{s3-cqq} through \eqref{s3-cmq} is terminated
at $5$\,ps, in evaluating the transport coefficients
listed in Table~\ref{table02}.
Onsager's reciprocal relation $M^{\mathrm{jm}}=M^{\mathrm{mj}}$
holds within the error of $6$\%, which gives a measure
of reliability of the numerical evaluation.

{
We note here that
for confined systems
the definition of the system volume
that is necessary in using Eqs.~\eqref{s3-cqq} through \eqref{s3-cmm}
 is not unique.
Here and in what follows, we use
the volume computed from the distance between 
the upper and lower wall atoms.
Another possibility could be 
to employ the domain actually occupied by the electrolyte solution
as the system volume.
The latter is smaller because of the excluded volumes of the wall atoms,
and using the latter yields difference in the transport coefficients
by several percent.
However, because the current density and mass flow density
obtained from the direct method also include
the system volume in the same manner
(${J}=j_x/V$ and ${Q}=c_x/V$), this difference
does not affect the relative comparison of the two methods.}

The current density 
${J}$ and mass flow density ${Q}$,
which are computed using the transport coefficients in Table~\ref{table02}
via Eq.~\eqref{s1-mat},
are valid only within the linear response regime.
In order to clarify quantitatively the range of the external field strength
in which Eq.~\eqref{s1-mat} is valid, 
we compare Eq.~\eqref{s1-mat} with the
 values of ${J}$ and ${Q}$ obtained with the direct method.
Figure~\ref{fig04} plots
${J}$ and ${Q}$ as functions of 
the external fields $E_x$ and $P_x$.
At each value of the external field strength,
a $4$\,ns production run is carried out to average over the time-series
data,
after a simulation
for $1$\,ns 
to reach the steady state.
{
In the parameter range of Fig.~\ref{fig04},
the fluxes in the direct method are
obtained with better statistics
than those in the simulations at thermal equilibrium, 
because the configuration of ions and water molecules
is perturbed more significantly by the external fields.
Therefore we used only one initial configuration
to evaluate the fluxes at each value of the external field strength.}

In Fig.~\ref{fig04},
Eq.~\eqref{s1-mat} with the values in Table~\ref{table02}
is also indicated by the solid line.
Although the results of the direct method are generally larger
than those obtained from  Eq.~\eqref{s1-mat},
they approach asymptotically as $E_x\to 0$ and $P_x\to 0$.
Particularly, the two results agree well 
in the range $E_x\le 0.02$\,V/{\AA} and $P_x\le 0.2$\,cal/g{\AA}.
The minimum values of the external fields in the figure are
$E_x=0.002$\,V/{\AA} and $P_x=0.02$\,cal/g{\AA},
which are extremely large values compared with
the field strength attainable in laboratories.
In  MD simulations,
very large external fields are usually necessary
to distinguish the observed variables from the thermal fluctuations
as in this case.
Figure~\ref{fig04} confirms that
even if the external field is unrealistically large,
there exists the range in which the results of the direct method
agree well with the results based on the linear response theory.
On the other hand, it implies that 
extrapolating a result of one single computation of the direct method
can cause serious errors when the external field is too strong,
as in the range $E_x>0.02$\,V/{\AA} of Figs.~\ref{fig04}(a) and (c), 
and in the range $P_x>0.3$\,cal/g{\AA} of Fig.~\ref{fig04}(b).

\begin{table*}[t]
\vspace*{0cm}
\vspace*{0cm}
\centering
\caption{
Transport coefficients;
values in parentheses are standard errors
for ten simulation runs.
}
\label{table02}
\vspace*{2mm}
\renewcommand{\arraystretch}{1.4}
\begin{tabular}{cccc}
\hline
\hline
 $M^{\mathrm{jj}}$\,[S/m] &  $M^{\mathrm{jm}}$\,[$10^{-6}$kg/Vms] &
 $M^{\mathrm{mj}}$\,[$10^{-6}$kg/Vms] &  $M^{\mathrm{mm}}$\,[$10^{-9}$kgs/m$^3$]\\
\hline
$6.52$ ($0.15$) & $24.0$ ($2.38$) & $22.5$ ($1.7$) & $1.47$ ($0.05$)\\
\hline
\hline
\end{tabular}
\end{table*}
\begin{table*}[t]
\begin{minipage}{0.7\textheight}
\centering
\caption{Parameters of the systems.}
\label{table03}
\vspace*{2mm}
\renewcommand{\arraystretch}{1.0}
\begin{tabular}{cccccccccc}
\hline
\hline
 & \multicolumn{5}{c}{system}\\
\cline{2-6}
 &$1$&  $2$&  $3$&  $4$&  $5$ \\
\hline
$\sigma$~[C/m$^2]$
 & $0.057$ & $0.082$ & $0.128$ &  $0.228$ & $0.514$\\ 
$N_x\times N_y$
 & $12\times 12$ & $10\times 10$ & $8\times 12$ &  $9\times 12$ & $8\times 12$\\ 
$\ell_c/\ell$
 & $6$ & $5$ & $4$ &  $3$ & $2$\\ 
no. of charged wall atoms
 & $16$ & $16$ & $24$ &  $48$ & $96$\\ 
no. of H$_2$O
 & $2840$ & $1980$ & $1890$ &  $2130$ & $1890$\\ 
no. of Na$^+$
 & $88$ & $66$ & $72$ &  $102$ & $144$\\ 
no. of Cl$^-$
 & $72$ & $50$ & $48$ &  $54$ & $48$\\ 
$H$~[{\AA}]
 & $42.3$ & $42.4$ & $41.9$ &  $41.2$ & $40.7$ \\
$C_{\mathrm{Na}}$~[M]
 & $1.54$ & $1.66$ & $1.91$ &  $2.44$ & $3.93$ \\
$C_{\mathrm{Cl}}$~[M]
 & $1.26$ & $1.26$ & $1.27$ &  $1.29$ & $1.31$ \\
$C_0$~[M]
 & $1.73$ & $1.74$ & $1.65$ &  $1.60$ & $1.55$ \\
\hline
\hline
\end{tabular}
\end{minipage}
\end{table*}
\begin{figure}[t]
\begin{center}
\includegraphics[scale=1.1]{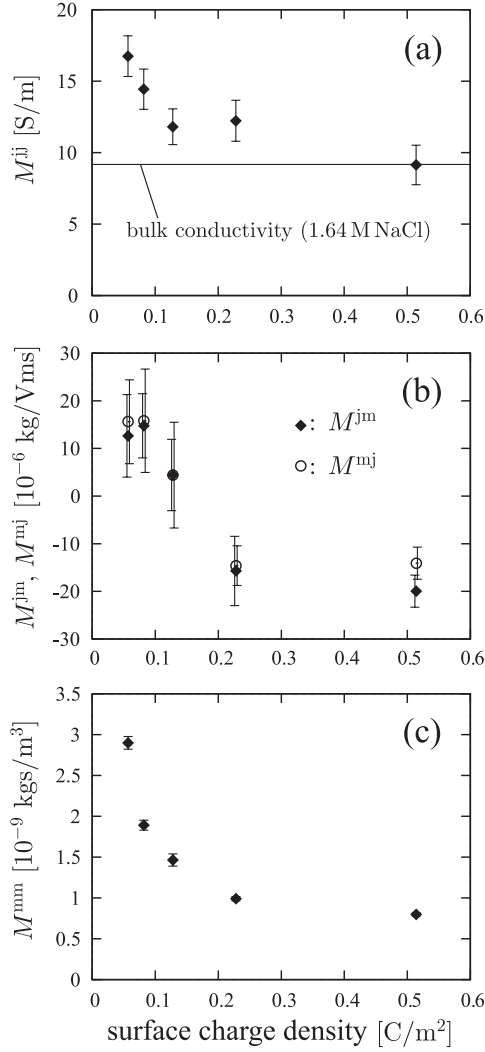} 
\caption{
Transport coefficients vs surface charge density
for the systems listed in Table~\ref{table03}.
The error bar indicates the standard error for ten simulation runs.
In panel (b), the error bars for $M^{\mathrm{jm}}$ and $M^{\mathrm{mj}}$ 
are slightly shifted to the left and right, respectively, for legibility.
}
\label{fig05}
\end{center}
\end{figure}  
%

%
\subsection{\label{sec_second}Influence of the surface charge density}

In this section, 
we consider five systems listed in Table~\ref{table03}
and investigate the influence of the surface charge density
on the electrokinetic flows in nanochannels.
Each wall consists of $N_x$ and $N_y$ unit cells
(Fig.~\ref{fig01}(a)) in the $x$- and
$y$-directions, respectively,
and a unit negative charge is assign to 
every $\ell_c/\ell$ atoms in the direction of the shortest distance of the
triangular lattice.
The surface charge density of system $3$ is identical to 
that of the system considered in the previous section,
though the concentration of ions in the present section
is higher.
The number of water molecules
contained between the gap is chosen
such that the distance $H$ is within $41.5\pm 1$\,{\AA}.
The number of Cl$^-$ ions is chosen
for the ion concentration to be within $1.3\pm 0.05$\,M,
and then the number of Na$^+$ is determined
from the electrical neutrality,
i.e., $C_{\mathrm{Na}}$
is increased as the surface charge density increases
to compensate the negative surface charge,
while $C_{\mathrm{Cl}}$ is kept at constant.

{
Since the concentration of ions is
sufficiently high in the systems considered herein,
the electrical double layers do not overlap,
and there exists a region near the center of the channel
where the profiles of concentration of Na$^+$ and Cl$^-$
exhibit plateaus with a common value.}
The values of the concentration of the plateau region,
denoted by $C_0$, are also listed in Table~\ref{table03}.

\begin{figure}[t]
\begin{center}
\includegraphics[scale=1.1]{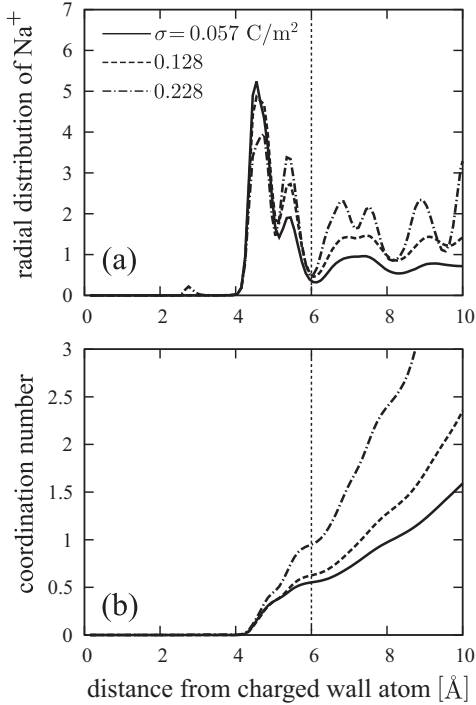} 
\caption{
(a) Radial distribution function (RDF) $g(r)$ of Na$^+$ about the charged wall
 atom. (b) Coordination number (or integrated RDF) $c_g(r)$.
}
\label{fig06}
\end{center}
\end{figure}

\begin{figure}[t]
\begin{center}
\includegraphics[scale=1.1]{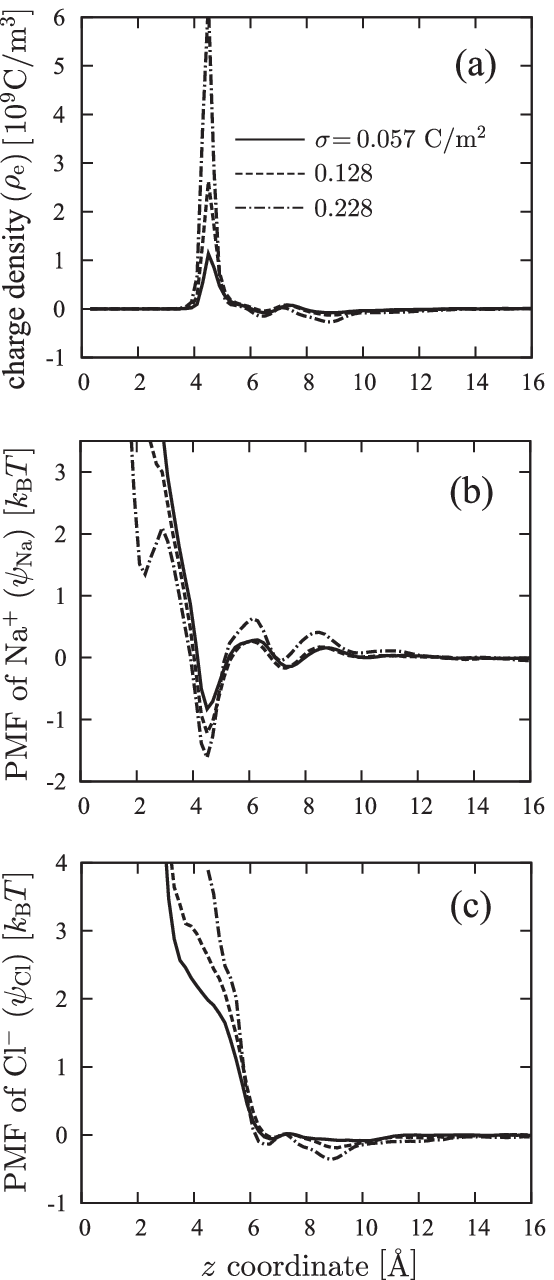} 
\caption{
Profiles in the $z$-direction of (a) the charge density, (b) the
 potential of mean force (PMF) for
 Na$^+$, and (c) that for Cl$^-$. The origin of the coordinate is at the position of the wall atoms.
}
\label{fig07}
\end{center}
\end{figure}

The transport coefficients obtained with the protocol described in the previous subsection
are plotted in Fig.~\ref{fig05}.
Here, the correlations for longer time interval ($0\le t\le 100$\,ps) 
than that in the previous subsection were taken in computing the
transport coefficients.
In order to interpret the results, 
the distribution of 
Na$^+$ is investigated.
{
We show in Fig.~\ref{fig06}
the radial distribution function $g(r)$ of Na$^+$ about the charged wall atom,
along with the coordination number defined as
$c_g(r)=2\pi n_0 \int_0^{r}\tilde{r}^2g(\tilde{r})\mathrm{d}\tilde{r}$
with $n_0$ being the average number density of Na$^+$.
Since the distribution of Na$^+$ is restricted to 
a half side of the wall atoms, 
the ions in a hemispherical shell
is counted in obtaining $g(r)$.}
In addition to the radial distribution of Na$^+$,
the distributions of
Na$^+$ and Cl$^-$ across the channel are also investigated.
For this purpose, the local concentrations
of ions, $C_{\mathrm{Na}}(z)$ and $C_{\mathrm{Cl}}(z)$, are
evaluated by counting  the ions
within $z\pm 0.1$\,{\AA} during the simulation for $5$\,ns.
Using the local concentration, the charge density distribution
$\rho_{\mathrm{e}}(z)$
and the PMFs $\psi_{\alpha}(z)$ ($\alpha=$Na, Cl) are calculated through the
following formulas~\cite{Chandler1987}
and plotted in Fig.~\ref{fig07}:
\begin{align}
&\rho_{\mathrm{e}}(z) = F(C_{\mathrm{Na}}(z)-C_{\mathrm{Cl}}(z)),
\label{s4-charge}\\
&\psi_{\alpha}(z)=-k_{\mathrm{B}}T\ln (C_{\alpha}(z)/C_0),
\label{s4-pmf}
\end{align}
where $F$ is the Faraday constant.


For comparison with the effective electrical conductivity
$M^{\mathrm{jj}}$ shown in Fig.~\ref{fig05}(a),
the electrical conductivity of the bulk NaCl solution at concentration
{
$1.64$\,M} 
is computed from Eq.~\eqref{s3-cqq} with
$\langle j_x(t)j_x(0)\rangle$ obtained from
a simulation with periodic boundary conditions in the three directions.
The electrical conductivity in the nanochannels
is generally higher than the bulk conductivity,
because of the surface conductivity.
The excess conductivity is reduced as
the surface charge density increases, in spite of the fact that
the number of excess counter-ions increases.
This is because
the counter-ions are strongly bound
due to the counter-ion condensation
near the surface, and thus the contribution to the conductivity
decreases. 
{
The behavior of the bound counter-ions is shown
in Fig.~\ref{fig06}:
the radial distribution
exhibits clear separation at $r=6$\,{\AA}, implying that
the counter-ions within $r<6$\,{\AA} are bound to a charged wall atom.
The coordination number at $r=6$\,{\AA}
is the number of bound counter-ions per charged wall atom.
It increases as the surface charge density increases,
indicating a reduced number of free counter-ions
for the high surface charge density.}

The transport coefficients for the streaming current
$M^{\mathrm{jm}}$ and the electro-osmotic flow $M^{\mathrm{mj}}$
are identical within the error, as shown in Fig.~\ref{fig05}(b),
which again confirms Onsager's reciprocal relation.
{
Note that they are negative 
at $\sigma=0.228$ and $0.514$\,C/m$^2$,
i.e., the direction of the streaming current and the electro-osmotic
flow are reversed.}
This is consistent with the reversal of the electro-osmotic flow
reported in Ref.~\onlinecite{QA2004}.
The cause of the inversion is understood as follows:
in the case of the large surface charge density,
the counter-ions are strongly bound at the well of PMF at $z=4.5$\,{\AA} (Fig.~\ref{fig07}(b)),
and the co-ions (Cl$^-$) are then pushed toward the middle of the channel.
The co-ions gather near the well of PMF at $z=9$\,{\AA} shown in
Fig.~\ref{fig07}(c), to form the negatively charged region
observed in Fig.~\ref{fig07}(a).
Responses of this negatively charged region to the mass acceleration and
electric field result in the reversed streaming current and
electro-osmotic flow.

{
We note here the contribution 
of the electro-osmotic flow to the surface conductivity
observed in Fig.~\ref{fig05}(a).
In the case of the forward flow, the
number of counter-ions in the mobile region is larger than that of
co-ions, and the electric current in the forward direction
is enhanced by the electro-osmosis.
On the other hand, when the electro-osmotic flow is reversed,
the number of co-ions exceeds that of counter-ions
in the mobile region,
increasing the speed of negative charge in the reverse direction.
The latter also contributes to the current in the forward
direction, or the conductivity gain.
Therefore, the main reason for the decreasing
conductivity with increasing surface charge density observed in Fig.~\ref{fig05}(a)
is most probably the loss of
mobile counter-ions due to the strong binding, as explained before.}

\begin{figure}[t]
\begin{center}
\includegraphics[scale=1]{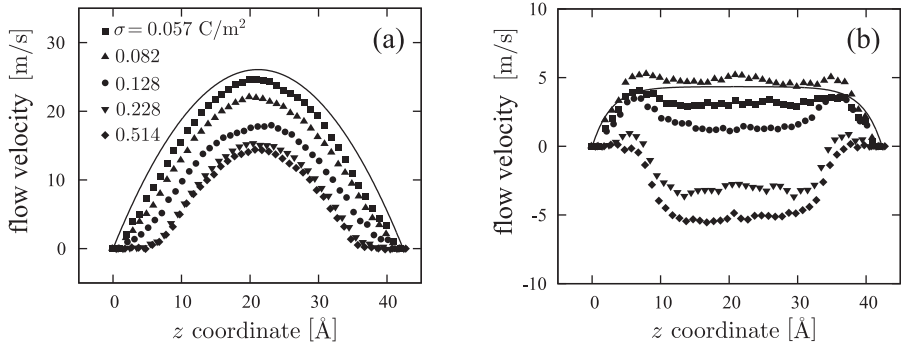} 
\caption{
Velocity profiles across the channel of 
(a) the Poiseuille-type flow at $P_x=0.2$\,cal/g{\AA} 
and (b) the electro-osmotic flow at $E_x=0.02$\,V/{\AA}. 
The solid line indicates the profiles predicted
by the continuum theory (Eq.~\eqref{s4-pb1});
$\sigma=0.057$\,C/m$^2$ in panel (b).
}
\label{fig08}
\end{center}
\end{figure}

Figure~\ref{fig05}(c) shows that the rate of the Poiseuille-type flow 
induced by the mass acceleration decreases as  the surface
charge density increases.
In order to investigate this flow reduction in greater details,
in Fig.~\ref{fig08}(a),
we show the velocity profiles 
obtained with the direct method at $P_x=0.2$\,cal/g{\AA}.
The electro-osmotic flows at
$E_x=0.02$\,V/{\AA} are also shown in Fig.~\ref{fig08}(b). 
The values of the field strength are within the
limit of the linear response regime (Fig.~\ref{fig04}).
The velocity at $z$ is the mean velocity of all the atoms
existing in the range $z\pm 0.5$\,{\AA},
during the $4$\,ns simulation.
For comparison, the velocity profiles
of the continuum theory
based on the Stokes equation and the Poisson--Boltzmann equation are
also shown:~\cite{YKW2014}
\begin{align}
&u_x(z)=\dfrac{P_x\rho_0}{2\mu}\left(zH-z^2\right)
\nonumber\\
&\qquad
       +\dfrac{E_x\sigma}{\mu\kappa}\left(\dfrac{\cosh(\kappa H/2)-\cosh(\kappa(z-H/2))}{\sinh(\kappa H/2)}\right),
\label{s4-pb1}\\
&\kappa=\left(\dfrac{2C_0 eF}{\varepsilon_0\varepsilon_r k_{\mathrm{B}}T}\right)^{1/2},
\label{s4-pb2}
\end{align}
where $\varepsilon_0$ is the permittivity of vacuum,
and $\mu$ and $\varepsilon_r$ are the viscosity and the dimensionless dielectric
constant, respectively, of the electrolyte solution.
%
{
For simplicity, 
we have assumed the stick boundary condition at the interface,
and the uniform viscosity 
($\mu=0.72\times 10^{-3}$\,Pa\,s)
and dielectric constant ($\varepsilon_r=76.7$) 
of the SPC/E water molecules,
which are listed in Table~II of Ref.~\onlinecite{WTV2006}.
An attempt to improve the continuum model
using the non-uniform viscosity and dielectric constant with slip
boundary condition is found in Ref.~\onlinecite{BN2012}.}
%
Clearly, from  Fig.~\ref{fig08}(a), the molecules
near the surfaces are immobile 
in the cases of large surface charge density.
{
The previous results
of the immobilization of the water molecules~\cite{SMC+2013}
and that of the counter-ion,~\cite{Netz2003}
for large surface charge densities,
are consistent with the present observation.}
Particularly, for $\sigma\ge 0.228$\,C/m$^2$, the flow velocity is
almost zero in $z\le 5$\,{\AA}.
The effective narrow gap due to the immobilization
of the molecules
results in a decrease of the transport coefficients, as shown
in Fig.~\ref{fig05}(c).

The velocity profiles of the electro-osmotic flow
are shown in Fig.~\ref{fig08}(b),
along with the continuum model at $\sigma=0.057$\,C/m$^2$.
Obviously, in Eq.~\eqref{s4-pb1},
the flow velocity of the continuum model increases in proportion to the
surface charge density.
On the contrary, the flow velocity obtained with the MD simulations
decreases as the surface charge density increases,
and the reversal of the flow
takes place for $\sigma\ge 0.228$\,C/m$^2$.
This behavior of the electro-osmotic flow is perfectly
consistent with the transport coefficients shown in Fig.~\ref{fig05}(b).
Although the driving force in the forward direction
acts on the positively charged region $z\le 5$\,{\AA} (Fig.~\ref{fig07}(a)),
the molecules in this region
do not move because of the strongly bound
counter-ions for $\sigma\ge 0.228$\,C/m$^2$, as in Fig.~\ref{fig08}(b).
Therefore, the force acting on the 
negatively charged region at $z=9$\,{\AA} (Fig.~\ref{fig07}(a)) drives
the flow in the opposite direction.

\begin{figure}[t]
\begin{center}
\includegraphics[scale=1.1]{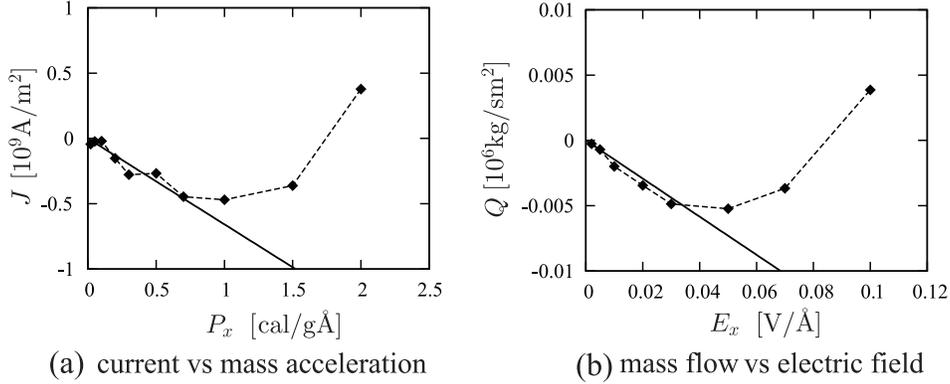} 
\caption{
(a) The current density induced by the mass acceleration, and
(b) the mass flow density induced by the electric field,
in the case of $\sigma=0.228$\,C/m$^2$.
See the caption of Fig.~\ref{fig04}.
}
\label{fig09}
\end{center}
\end{figure}  
\begin{figure}[t]
\begin{center}
\includegraphics[scale=1.1]{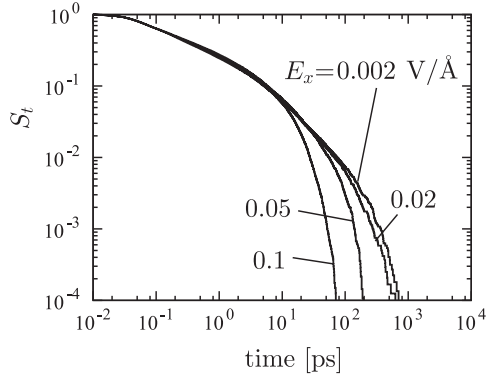} 
\caption{
Bond survival probability distribution $S_t$ 
for a bond between an Na$^+$ ion and a negatively charged wall atom
in the presence of the external electric field.
The surface charge density is $\sigma=0.228$\,C/m$^2$.
}
\label{fig10}
\end{center}
\end{figure}  

The observation above implies that the 
counter-ion condensation takes place
for $\sigma\ge 0.228$\,C/m$^2$,
in view of the fact that 
the mobility of the counter-ions condensed at the interface
is significantly lower than that of the 
weakly bound
counter-ions.~\cite{WK2006}
The recent counter-ion condensation theory
for plane surfaces by Manning
predicts the critical value above which the counter-ions condense at the
interface:~\cite{Manning2010}
\begin{equation}
\sigma_{\mathrm{crit}}=\dfrac{2\varepsilon_0\varepsilon_r k_{\mathrm{B}}T\kappa(-\ln\kappa
 l_{\mathrm{ref}})}{e^2},
\label{s4-crit}
\end{equation}
where $l_{\mathrm{ref}}$ is  the characteristic length 
assumed to be small compared with the thickness of the electrical double layer;
the possibility of identifying $l_{\mathrm{ref}}$ with the length scale
of the molecular structure at the surface is discussed in Ref.~\onlinecite{Manning2010}.
If we apply Eq.~\eqref{s4-crit} to our case with assuming $l_{\mathrm{ref}}=1$~{\AA},
then the predicted critical value is $\sigma_{\mathrm{crit}}= 0.135$\,C/m$^2$.
Although the theory in Ref.~\onlinecite{Manning2010} treats
perfectly plane surface and states no dynamical property 
in the direction parallel to the surfaces,
the consistency with the present simulation results 
($\sigma_{\mathrm{crit}}=0.128\sim 0.228$\,C/m$^2$),
in conjunction with the previous results of the mobility of the
condensed counter-ions in another geometry,~\cite{WK2006}
could shed light on the interplay between the
counter-ion condensation and dynamical properties of the counter-ions 
adjacent to realistic plane surfaces. 

We now examine the
influence of the external field strength
on the streaming current and electro-osmotic flow,
which are reversed in the linear response regime.
In Fig.~\ref{fig09}, 
we plot ${J}$ as a function of $P_x$,
and ${Q}$ as a function of $E_x$, for the case of $\sigma=0.228$\,C/m$^2$.
Similarly to Fig.~\ref{fig04}, $J$ and $Q$ asymptotically approach the
results of the linear response theory as $E_x \rightarrow 0$ and $P_x\rightarrow 0$.
They start to depart at $P_x=1$\,cal/g{\AA} and $E_x=0.05$\,V/{\AA},
and the current and the mass flow change
the direction at $P_x=2$\,cal/g{\AA} and $E_x=0.1$\,V/{\AA}, respectively,
because the bound counter-ions are pulled away from the
surface charges by the strong external fields.
The motion of the bound counter-ions is described
in a quantitative manner using the bond survival probability distribution:
${S}_t(t)=1-\int_0^{t}{P}_t(s)\mathrm{d}s$ 
with ${P}_t(s)\mathrm{d}s$ being the probability that a 
counter-ion stays within $6$\,{\AA} form a charged wall atom
for time period $s$. 
{
Recall that the counter-ions within $6$\,{\AA} are bound (see
Fig.~\ref{fig06}).}
The bond survival probability distribution
is shown in Fig.~\ref{fig10} for some values
of the external electric field at $\sigma=0.228$\,C/m$^2$.
Clearly, the strong external electric field
shortens the bond survival time,
meaning that the counter-ions are dragged by the field.

Figure~\ref{fig09} clearly demonstrates that the property of the flows
for the same system can drastically differ depending on the
external field strength, implying that we should be careful in
extrapolating the results of the direct method to realistic systems.

\begin{table*}[t]
\begin{minipage}{0.7\textheight}
\centering
\caption{Parameters of the systems.}
\label{table04}
\vspace*{2mm}
\renewcommand{\arraystretch}{1.0}
\begin{tabular}{cccccccccc}
\hline
\hline
 & \multicolumn{4}{c}{system}\\
\cline{2-5}
 &$1'$ &  $2'$ &  $3'$ &  $4'$ \\
\hline
$\sigma$~[C/m$^2$]
 & $0.057$ & $0.082$ & $0.128$ &  $0.228$ \\ 
no. of H$_2$O
 & $2840$ & $1980$ & $1890$ &  $2130$ \\ 
no. of Na$^+$
 & $74$ & $52$ & $49$ &  $54$ \\ 
no. of Cl$^-$
 & $58$ & $36$ & $25$ &  $6$ \\ 
$H$~[{\AA}]
 & $42.1$ & $42.1$ & $41.4$ &  $40.1$ \\
$C_{\mathrm{Na}}$~[M]
 & $1.29$ & $1.31$ & $1.30$ &  $1.29$ \\
$C_{\mathrm{Cl}}$~[M]
 & $1.01$ & $0.90$ & $0.66$ &  $0.14$ \\
$C_0$~[M]
 & $1.55$ & $1.17$ & $1.08$ &  $0.15$ \\
\hline
\hline
\end{tabular}
\end{minipage}
\end{table*}
\begin{figure}[t]
\begin{center}
\includegraphics[scale=1.1]{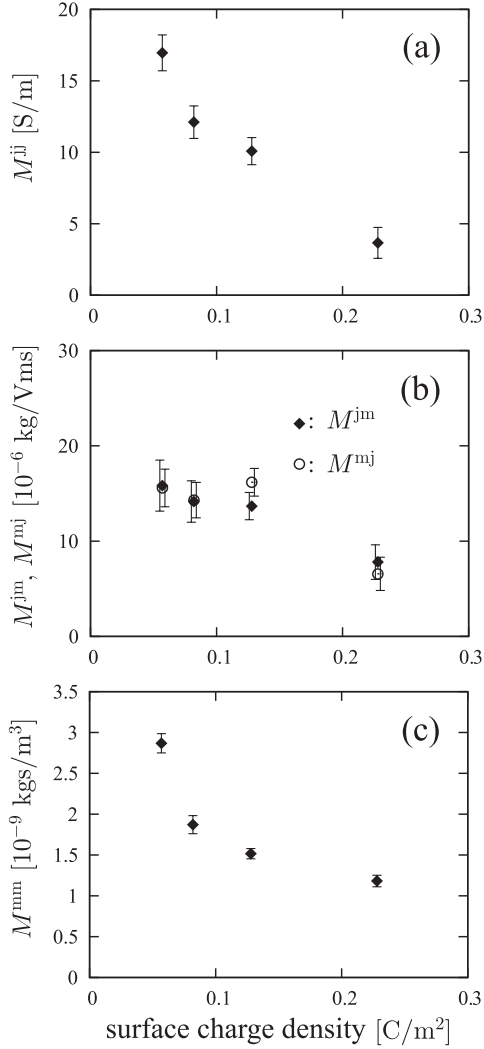} 
\caption{
Transport coefficients vs surface charge density
for the systems listed in Table~\ref{table04}.
See the caption of Fig.~\ref{fig05}.
}
\label{fig11}
\end{center}
\end{figure}  

We conclude with a brief discussion on the concentration of the solution in the
charged nanochannel.
In the simulations for the systems listed in Table~\ref{table03},
the counter-ions (Na$^+$ ions) were added
to compensate the increasing surface charge density.
We here explore the other possibility of compensating
the increasing negative charge on the surface, namely,
decreasing the concentration of the co-ions (Cl$^-$ ions) while
maintaining the concentration of the counter-ions (Na$^+$ ions) at constant.
Table~\ref{table04} lists the simulation parameters used here.
The configurations of the wall atoms in systems $1'\sim 4'$ are
exactly the same as those of systems $1\sim 4$ in Table~\ref{table03}, respectively.
The number of Na$^+$ ions is chosen
such that the concentration falls within $1.3\pm 0.05$\,M,
and then the number of Cl$^-$ is determined
from the charge neutrality. 

The transport coefficients
of systems $1'\sim 4'$ are shown in Fig.~\ref{fig11}.
One obvious qualitative difference from Fig.~\ref{fig05}
is the significant decrease of the conductivity at 
$\sigma=0.228$\,C/m$^2$ in Fig.~\ref{fig11}(a).
This decrease  is caused by the fact that most of the counter-ions are
condensed at the interface in the case of the high surface charge
density, and the number of the mobile ions are greatly reduced.
Indeed, the value of $C_0$, which represents the concentration
in the mobile region, of system~$4'$ is very small compared with that of system~$4$.
Another important difference is that
the reversal of the streaming current and the electro-osmotic flow
does not occur at $\sigma= 0.228$\,C/m$^2$ (Fig.~\ref{fig11}(b)),
i.e., $M^{\mathrm{jm}}$ and $M^{\mathrm{mj}}$ do not change their sign,
although the values are slightly 
smaller than those for $\sigma< 0.228$\,C/m$^2$.
This is because the number of co-ions is not sufficient to form the negatively charged
region observed in Fig.~\ref{fig07}(a).

If an experiment corresponding to the systems considered in the present study
is performed for a setup with two reservoirs of constant concentration $C_s$ 
connected by a nanochannel,  
$C_0$ will be comparable  to $C_s$. Therefore,
the systems listed in Table~\ref{table03}
are relevant because the value of $C_0$ is controlled
such that it ranges within $1.65\pm 0.1$\,M, whereas
the value of $C_0$ varies significantly in Table~\ref{table04}.
It is then important to notice that the value of $C_0$ that is common to Na$^+$
and Cl$^-$ is defined only if the thickness of the electrical double layer
is sufficiently short compared with the channel width,
as in the systems considered in the present paper; 
otherwise the concentration of Na$^+$ differs from that of Cl$^-$ 
over the channel. In the latter case
the Donnan effect manifests itself~\footnote{Note that according to
Ref.~\onlinecite{BC2010B} Donnan effects could appear as soon as the channel
width becomes comparable to the Dukhin length $\sigma/FC_0$; in our case
this length is indeed comparable to the width for the highest charge
densities, which highlights the importance of considering the
possibility of a reduction in free salt concentration. } and 
the concentration of Cl$^-$ should decrease upon
increasing the surface charge density (see, e.g., Ref.~\onlinecite{BC2010B}). 
Therefore care must be taken in setting the value of
concentration for systems with different surface charge densities,
depending on the circumstances,
because the manner of changing the concentration has 
significant influence on the qualitative behavior of the
electrolyte solution, 
as demonstrated by comparing Figs.~\ref{fig05} and \ref{fig11}.
%
%
%
\section{Summary}

In the present paper, we have studied
the currents and mass flows 
of an aqueous NaCl solution in nanochannels of the gap~$\sim 40$\,{\AA}
induced by an electric field and a mass acceleration corresponding to a pressure gradient.
In order 
to accurately calculate the four transport coefficients 
through the Green--Kubo formulas,
ten MD simulation runs with different initial configurations
for each system are carried out to obtain 
smooth time-correlation functions.
Comparison of the current and mass flow
predicted by Eq.~\eqref{s1-mat}
with those obtained by the direct method
revealed that,
although extremely strong external fields
led to large discrepancies, 
the two results converged 
within a range of  external field strengths
for which the flows were still distinguishable from the thermal fluctuation.
In the present study, we considered the time-independent external fields.
However, the responses to the time-dependent fields, such as oscillatory
fields, can be also examined by means of both the Green-Kubo formulas and the
direct method, the results of which should coincide in the linear
response regime.~\cite{Mizuno2012}

{
One of the advantages of the direct method using nonequilibrium
simulation is that the
flow induced in the channel is obtained 
with better statistics than those in the simulation at thermal equilibrium, 
because of significant perturbation due to the external fields.
Therefore, generally the computational cost 
required to obtain the flow at a specified field strength is much less than
that required to perform the time-integration in the Green--Kubo formulas,
as discussed in Section~\ref{sec_first}.
However, extrapolating only a few results of the direct method
can mislead us, as demonstrated in Fig.~\ref{fig09}. 
A careful examination is inevitable to ascertain if the 
field strength is in the linear or non-linear regime,
the cost of which can comparable with that for the Green--Kubo approach.
Therefore, in studies on responses to external fields,
it is preferable to first investigate the system properties through the Green--Kubo approach,
followed by the additional computations to observe the flow directly,
as has been done in Fig.~\ref{fig08}, or to investigate the non-linear
regime as necessary.
A relevant discussion is found in a recent note
on the computation of the bulk viscosity in Ref.~\onlinecite{CSB2009}.}

The influence of the surface charge density
of the channel walls was also examined,
with maintaining the channel width 
and the concentration of Cl$^-$ at constant.
As a result,
the effective electric conductivity
and the rate of the Poiseuille-type flow
were found to be reduced by the large surface charge density,
because the excess Na$^+$ ions
strongly bound near the interface
interfered with the
charge and mass flows.
The reversal of the streaming current and the electro-osmotic flow
was observed both in the 
transport coefficients obtained with the Green--Kubo formulas and in the
results of the direct method, which is consistent with the
finding reported in Ref.~\onlinecite{QA2004}.

As an extension of the present study,
it would be interesting to investigate the influence of the variety of the surfaces,
for example, hydrophobic surfaces and more complicated chemically modified surfaces.
Another direction of future studies could be 
to replace the solute and solvent by 
the more complex ones used in lithium ion batteries and fuel cells 
for understanding the nano-scale transport properties important in 
the state-of-the-art electrochemical devices. Our study shows that,
apart from 
{
the computational cost in dealing with more complex systems (for which nonlinear effects may be even more important than shown here)   and}
the difficulty in identifying the appropriate force fields,
there is no principle difficulty in obtaining accurate values of the
electro-osmotic coefficients at this scale using molecular dynamics.


\begin{acknowledgments}
The authors are grateful to S. Iwai for computer assistance in preparing the manuscript.
H.~Y., T.~K., and H.~W. are supported by MEXT program ``Elements
Strategy Initiative to Form Core Research Center'' (since 2012).  (MEXT
stands for Ministry of Education, Culture, Sports, Science, and Technology, Japan.)
H.~M. acknowledges support by the Nanosciences Foundation of Grenoble.
J.-L.~B. is supported by the Institut Universitaire de France, and acknowledges useful discussions with L. Bocquet and E. Charlaix.
\end{acknowledgments}

%
%
%
\end{document}